\documentclass[prl,twocolumn,superscriptaddress]{revtex4-1}
\usepackage{braket}
\usepackage{upgreek}
\usepackage{}
\usepackage{amssymb}
\usepackage{amsfonts}
\usepackage{bbm}
\usepackage{mathrsfs}
\usepackage{graphics,graphicx,epsfig,bm,amsmath,amsthm,amssymb}
\usepackage{bm}
\usepackage{bbm}
\usepackage{longtable}
\usepackage{multirow}
\usepackage{array}
\usepackage{color}
\usepackage{diagbox}
\usepackage[usenames,dvipsnames]{xcolor}

\usepackage{float}
\usepackage[a4paper,colorlinks=true,
linkcolor=blue,citecolor=blue,
pdfauthor={ },
pdftitle={ },
pdfsubject={ },
pdfkeywords={ }]{hyperref}

\begin{document}

\title{Observation of the Knot Topology of Non-Hermitian Systems in a Single Spin}

\author{Yang Wu}
\author{Yunhan Wang}
\author{Xiangyu Ye}
\author{Wenquan Liu}
\affiliation{CAS Key Laboratory of Microscale Magnetic Resonance and School of Physical Sciences, University of Science and Technology of China, Hefei 230026, China}
\affiliation{CAS Center for Excellence in Quantum Information and Quantum Physics, University of Science and Technology of China, Hefei 230026, China}

\author{Chang-Kui Duan}
\author{Ya Wang}
\author{Xing Rong}
\email{xrong@ustc.edu.cn}
\affiliation{CAS Key Laboratory of Microscale Magnetic Resonance and School of Physical Sciences, University of Science and Technology of China, Hefei 230026, China}
\affiliation{CAS Center for Excellence in Quantum Information and Quantum Physics, University of Science and Technology of China, Hefei 230026, China}
\affiliation{Hefei National Laboratory, University of Science and Technology of China, Hefei 230088, China}
\author{Jiangfeng Du}
\email{djf@ustc.edu.cn}
\affiliation{CAS Key Laboratory of Microscale Magnetic Resonance and School of Physical Sciences, University of Science and Technology of China, Hefei 230026, China}
\affiliation{CAS Center for Excellence in Quantum Information and Quantum Physics, University of Science and Technology of China, Hefei 230026, China}
\affiliation{Hefei National Laboratory, University of Science and Technology of China, Hefei 230088, China}
\affiliation{School of Physics, Zhejiang University, Hangzhou 310027, China}

\begin{abstract}                                   
The non-Hermiticity of the system gives rise to distinct knot topology that has no Hermitian counterpart.
Here, we report a comprehensive study of the knot topology in gapped non-Hermitian systems based on the universal dilation method with a long coherence time nitrogen-vacancy center in a $^{\text{12}}$C isotope purified diamond.
Both the braiding patterns of energy bands and the eigenstate topology are revealed. 
Furthermore, the global biorthogonal Berry phase related to the eigenstate topology has been successfully observed, which identifies the topological invariance for the non-Hermitian system.
Our method paves the way for further exploration of the interplay among band braiding, eigenstate topology and symmetries in non-Hermitian quantum systems.
\\
\end{abstract}

\maketitle

The non-Hermitian (NH) Hamiltonian has drawn intensive attention \cite{RMP_Bergholtz,PRL_NHtopo1,PRX_NHtopo1,PRX_NHtopo2,Science_NHtopo,PRL_NHtopo2,PRA_NHtopo} due to its application in photonic \cite{Science_NHApli1,NP_NHApli,Science_NHApli2,Science_NHApli3,Nature_PhotoNH,Science_PhotoNH1,PRL_PhotoNH1,PRL_PhotoNH2,PRL_PhotoNH3,NaturePhotonics_PhotoNH,NP_PhotoNH} and acoustic \cite{NHAcous_Romain,NHAcous_Tuo,NHAcous_Zhi} systems with gain and loss, open quantum systems \cite{JPA_OpenNH,Nature_OpenNH,PRL_OpenNHCar,NP_OpenNHDiehl} and condensed matter systems of quasi-particles with finite lifetime \cite{Arxiv_CondNH,PRB_CondNH1,PRB_CondNH2,PRL_CondNH}. In contrast to Hermitian systems, topological structures that are exclusive to NH systems give birth to many intriguing phenomena like NH bulk boundary correspondence \cite{PRL_Kai,PRL_Dan} and NH skin effects \cite{PRL_Nobuyuki}. The emergence of NH nodal phases and gapped phases as well as their interplay with symmetry also adds new ingredients for NH topological band theory \cite{RMP_Bergholtz}. Understanding the topology introduced by NH physics helps one to explore topologically robust quantities and non-trivial edge states. All these uncommon phenomena provide insights into NH systems.

More recently, researchers have found a framework based on homotopy theory \cite{PRB_Fan,PRL_Haiping,PRB_ZhiLi} to classify the non-Hermitian topological phases.
Compared with the K-theory \cite{PRX_NHtopo1}, this classification method does not need to assume specific symmetries and can reveal more topological invariants.
To be precise, the Hamiltonian can be regarded as a mapping from the Brillouin zone to the space of energy bands and eigenstates. 
Thus the classification is done by finding all topologically non-equivalent mappings. 
Each class of mapping corresponds to a topological phase with a specific topological invariant.
The non-Hermiticity of the Hamiltonian can generate complex eigenvalues, which can constitute the knot structures.
The whole classification set of NH systems with knot topology is decomposed into several sectors, based on the braiding of eigenvalues,
and each sector can be further classified with the eigenstate topology \cite{PRB_ZhiLi}.

There are several experimental studies recently aiming at revealing the knot topology of the NH systems. 
For example, the braiding structure of energy bands has been observed in optical \cite{Nature_Fan}, mechanical \cite{Nature_Harris} and trapped ion \cite{PRL_Cao} systems. 
A comprehensive demonstration of the knot topology, especially the essential information from the eigenstate topology requires coherent evolution under the NH Hamiltonian. 
It is difficult to realize NH Hamiltonians in quantum systems since closed systems are generally governed by Hermitian Hamiltonians. 
Apart from the difficulty in realizing a NH Hamiltonian, the inevitable decoherence of the quantum system tends to destroy the coherent evolution. 
In order to overcome these difficulties, we applied the universal dilation method \cite{Science_YW,PRL_Wengang,PRL_Uwe,PRL_Kohei} to realize the NH Hamiltonians and fabricated a 99.999\% $^{\text{12}}$C isotope purified sample in which the decoherence had been significantly suppressed. 
Taking the one-dimensional (1D) NH lattice model as an example, we experimentally studied the knot topology in the NH Hamiltonian by measuring not only its energy eigenvalues, but also the corresponding eigenstates to reveal the global Berry phase. 
Compared with the NH topology shown in previous work \cite{PRL_Wengang} on the NV center, which stems from encircling varying numbers of exceptional points, the knot topology investigated in our work arises from both the braiding of all energy bands and the eigenstate topology.
Our platform is capable of dealing with other models and our results shed light on the complexity of topological features aroused by non-Hermiticity.

\begin{figure}[http]
	
	\centering
	
	\includegraphics[width=1.0\columnwidth]{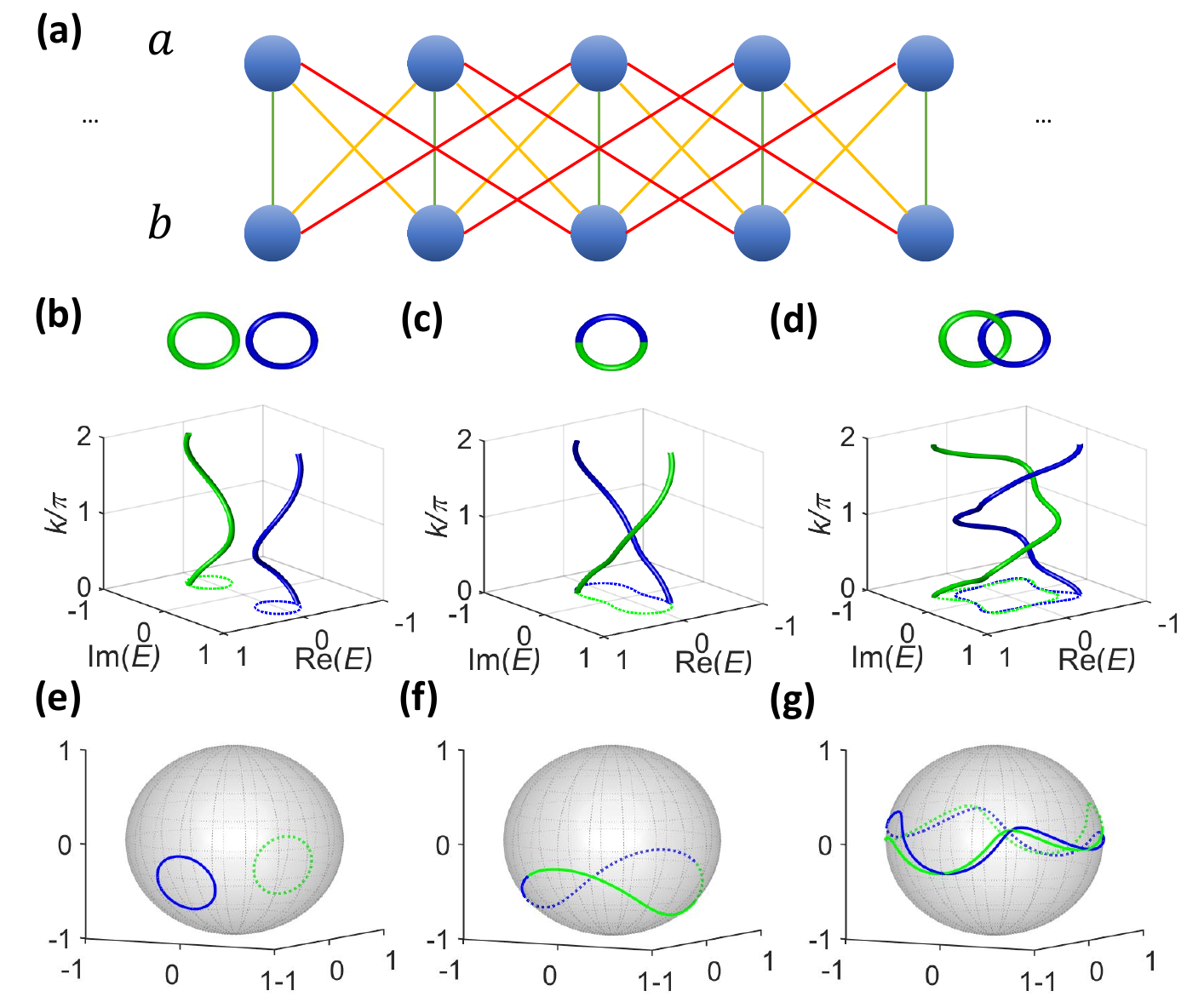}
	\caption{The knot topology of eigenvalues and eigenstates in a 1D NH lattice model. (a) Part of the lattice with on-site (green), nearest (yellow) and next nearest (red) interaction. (b-d) Band structure and the corresponding knot diagrams of three phases under different parameter configurations. The projection of eigenvalues on the complex plane (dashed lines) shows different topological structures. (b) Two completely separated bands, which are similar to an insulator in the Hermitian case. (c) Two bands exchange and the projection lines merge into a whole circle. (d) Two bands encircle each other and become the Hopf link. (e-g) The eigenstates show the same pattern as the corresponding energy bands.
	}
	\label{Fig1}
\end{figure}

Here we consider a two-band NH 1D lattice model (Fig.~\ref{Fig1}(a)) whose Hamiltonian may be written as
\begin{equation}
	\begin{split}
	&\mathcal{H}^{(m)} = \sum_{j} [ \Gamma^{0+}a_j b_j^\dagger + \Gamma^{0-}a_j^\dagger b_j +\\
	&\sum_{n=1}^{m} (\Gamma^{n+}_{1}a_jb_{j+n}^\dagger + \Gamma^{n-}_{1}a_j^\dagger b_{j+n} + \Gamma^{n+}_{2}a_{j+n}^\dagger b_j +\Gamma_{2}^{n-}a_{j+n}b_j^\dagger) ] ,
	\end{split}
\end{equation}
where $a_j\ (b_j)$ is the annihilation operator of sublattice $a\ (b)$ on the $j$th site, and the $\Gamma$'s are complex numbers representing the coupling strengths, i.e., $\Gamma^{0\pm}$ is the on-site interaction and $\Gamma_{1(2)}^{n\pm}$ is the hopping between sublattice $a\ (b)$ at site $j$ and lattice $b\ (a)$ at site $j+n$. Under the periodic boundary condition, the corresponding momentum space Hamiltonian is
\begin{equation}
	\begin{aligned}
		&H^{(m)}(k) = \left[\begin{array}{ccc}
			0 & \Gamma^{0-}\\
			\Gamma^{0+} & 0
		\end{array}\right] + \\
		&\sum_{n=1}^{m}\left[\begin{array}{ccc}
			0 & \Gamma_{1}^{n-}e^{ink}+\Gamma_{2}^{n+}e^{-ink}\\
			\Gamma_{1}^{n+}e^{-ink}+\Gamma_{2}^{n-}e^{ink} & 0
		\end{array}\right].
	\end{aligned}
\end{equation}
This model holds plenty of topological phases under different parameter configurations that ensure bands being separable. 
For $m=1$, i.e., the nearest coupling case, the system may be in the unlink phase or the unknot phase. Because $k=0$ and $ 2\pi$ are equivalent due to the periodic condition, the energy bands form closed curves. 
When the parameters are set as $\Gamma^{0-} = -0.45,$ $\Gamma^{0+}=0.79,$ $\Gamma^{1-}_{1}=-0.30i,$ $\Gamma^{1-}_{2}=0.08i,$ and $\Gamma^{1+}_{1,2}=0$, the non-Hermitian system is in the unlink phase, where the two circles are unrelated, as shown by Fig.~\ref{Fig1}(b).
If the parameters are changed to $\Gamma^{0-} = -0.21,$ and $\Gamma^{0+}=0.70$, the system will be in the unknot phase. 
There is only one circle because the two bands exchange (Fig.~\ref{Fig1}(c)). 
For $m=2$, the next nearest coupling enriches the topological phases in addition to the unlink and unknot. 
When the parameters are set as $\Gamma^{0-} = 0.04,$ $\Gamma^{0+}=0.49,$ $\Gamma^{1-}_{1}=\Gamma^{1+}_{2}=-0.13i,$ $\Gamma^{1-}_{2}=\Gamma^{1+}_{1}=0.02i,$ $\Gamma^{2-}_{1}=-0.58i,$ $\Gamma^{2+}_{2}=-0.21i,$ $\Gamma^{2+}_{1}=0.03i,$ and $\Gamma^{2-}_{2}=0.09i$, the two eigenvalues form a non-trivial braiding structure.
The energy bands braid around each other exactly once and this gives the Hopf link phase (Fig.~\ref{Fig1}(d)). 
The eigenstates also exhibit behaviors similar to those of the corresponding energy bands as shown in Fig.~\ref{Fig1}(e), \ref{Fig1}(f) and \ref{Fig1}(g). For larger $m$, the energy bands may braid more times, which leads to more phases. 
The classification based on homotopy theory has shown that all phases of this model are described by the braiding group $B(2)$ \cite{PRB_Fan,PRB_ZhiLi}.

We now study the topological phases of our model and the corresponding topological invariants in a quantum system based on the eigenvalues and eigenstates. 
Both the eigenvalues and the eigenstates can be obtained from the evolution under the NH Hamiltonian $H_s(k) = H^{(m)}(k)$. 
Let $\ket{\psi_{1,2}(k)}$ (the $k$ dependence is omitted for simplicity in the following) be the eigenstates with complex eigenvalues $E_{1,2} = E^r_{1,2}+iE^i_{1,2}$, where $E^r_n$ and $E^i_n$ are real and imaginary parts of $E_n$, respectively. 
For any initial state written as $\ket{\psi(0)}=c_1\ket{\psi_1}+c_2\ket{\psi_2}$, the evolution governed by $H_s(k)$ gives $\ket{\psi(t)}\propto c_1 e^{-iE^r_1t+E^i_1t}\ket{\psi_1}+c_2 e^{-iE^r_2t+E^i_2t}\ket{\psi_2}$.
Without loss of generality, we assume $E^i_1>E^i_2$. 
The eigenvalues can be extracted from the time evolution of the populations of $\ket{\psi_{1,2}}$, and the steady state of NH evolution will be the eigenstate $\ket{\psi_1}$ since $E^i_1$ is larger than $E^i_2$. 
By implementing evolution governed by $-H_s$, the eigenstate $\ket{\psi_2}$ can be obtained, which is due to $-E^i_2>-E^i_1$ for $-H_s$.

The evolution under the NH Hamiltonian $H_s(k)$ can be realized based on the universal dilation method. 
The state of the system $\ket{\psi(t)}$ needs to evolve as $i\partial_t\ket{\psi(t)}=H_s(k)\ket{\psi(t)}$. 
By introducing an ancilla, the NH evolution can be realized in a subspace while the total Hamiltonian $H_{\rm tot}$ is Hermitian. 
The initial state that reads $\ket{0}_s\ket{-}_a + \eta(0)\ket{0}_s\ket{+}_a$ evolves to $\ket{\psi(t)}_s\ket{-}_a+\eta(t)\ket{\psi(t)}_s\ket{+}_a$ under the Hamiltonian $H_{\rm tot}$, where $\ket{\pm}$ are eigenstates of $\sigma_y$ and $\eta(t)$ is a properly chosen time-dependent operator. 
Thus in the $\ket{-}_a$ subspace, apart from a normalization constant, the evolution is strictly governed by $H_s(k)$.

\begin{figure}[http]
	
	\centering
	
	\includegraphics[width=1\columnwidth]{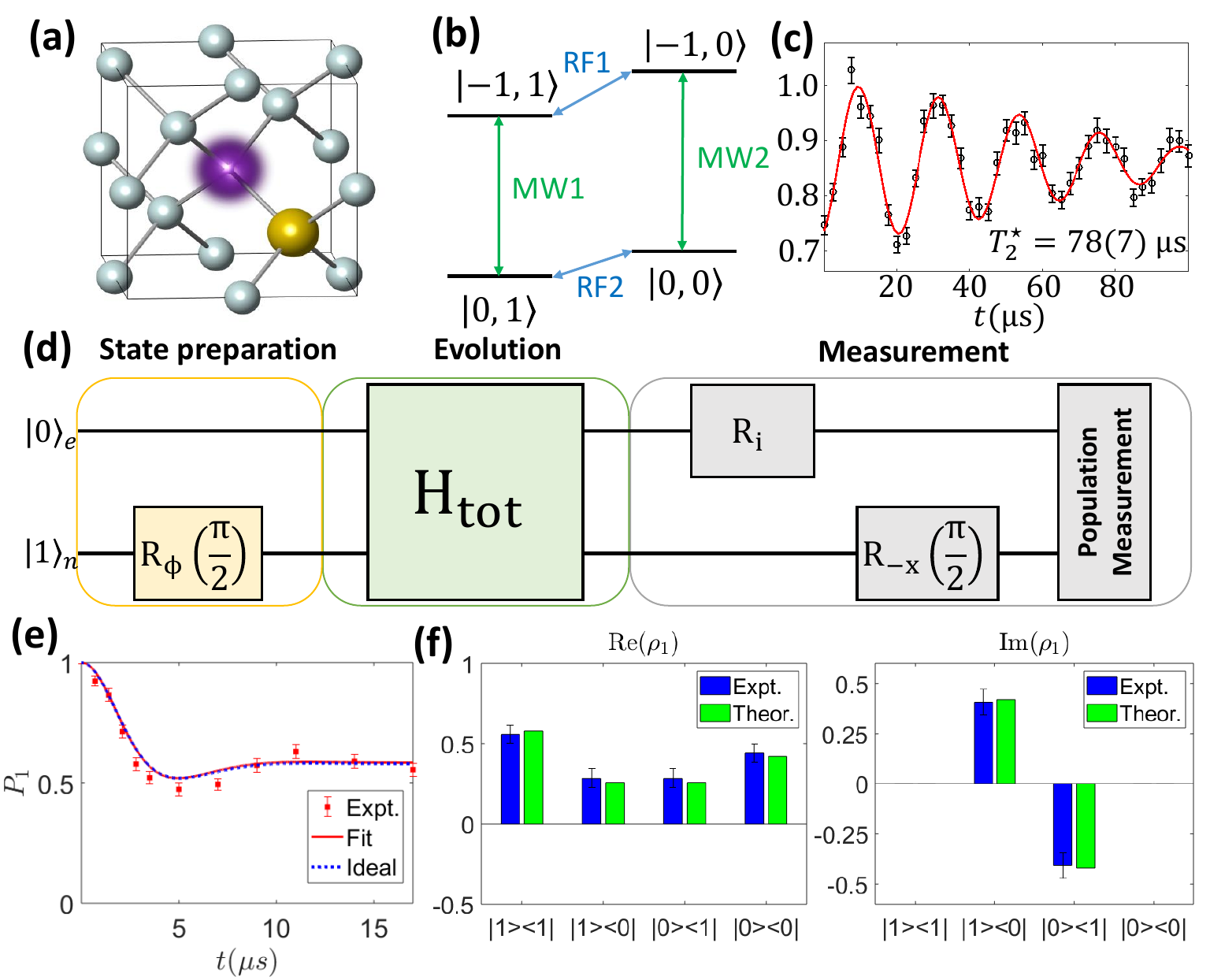}
	
	\caption{Realization of the NH Hamiltonian in the NV center. (a) Simplified atomic structure of an NV center. Light cyan balls are carbon atoms, the yellow ball is a nitrogen atom and the purple one is a vacancy. (b) Four used energy levels of the NV ground state with hyperfine and nuclear quadrupolar interaction. Microwave pulses with different frequencies can be applied to selectively drive the electron spin in the $\ket{m_I=1,0}$ subspace. Similarly, the radio-frequency (RF) field is used to selectively control the nuclear spin. (c) Dephasing time of the NV center in our experiment. Fitting of luminescence in the Ramsey experiment shows that $T_2^\star = 78(7)\ \upmu$s, which is long enough to maintain coherence in our experiment. (d) The pulse sequence of our experiment. State preparation is realized by laser pumping and an RF $\pi/2$ pulse $R_{\phi}(\pi/2)$ with the phase $\phi$ depending on $\eta_0$. The evolution under $H_{tot}$ is realized by two selective MW pulses with time-dependent amplitude, frequency and phase. Experimental results for the Hopf link phase; $k=0.6\pi$ are shown in panels (e) and (f). (e) Dynamics of the population. (f) Real and imaginary parts of the MLE results for the final state of evolution. Orange bars are theoretical predictions and blue bars with error bars are experimental results. 
	}
	\label{Fig2}
\end{figure}

We use a single nitrogen-vacancy (NV) center in diamond to experimentally realize the momentum space NH Hamiltonian for different phases with $k\in[0,2\pi]$ (see Appendix A for details of the experimental setup).
The NV center is a type of point defect in diamond that is composed of a nitrogen atom and a neighbor vacancy as shown in Fig.~\ref{Fig2}(a). 
The ground-state of the NV center is a triplet state with an electronic spin $S=1$ that interacts with the nuclear spin $I=1$ of the substitutional $^{\text{14}}$N.
We construct the dilated Hamiltonian $H_{\rm tot}$ in the subspace spanned by the states $\ket{m_S,m_I}=\ket{0,1},\ket{0,0},\ket{-1,1}$, and $\ket{-1,0}$ as shown in Fig.~\ref{Fig2}(b).
The energy levels are relabeled as $\ket {1}_e\ket{1}_n,\ket{1}_e\ket{0}_n,\ket{0}_e\ket{1}_n$, and $\ket{0}_e\ket{0}_n$, respectively.
The electron spin is chosen to be the system and nuclear spin serves as the ancilla.

The pulse sequence used to realize the evolution under the NH Hamiltonian is shown in Fig.~\ref{Fig2}(d). 
The external static magnetic field was set to be 506 G. 
The state of the NV center was polarized to $\ket{0}_e\ket{1}_n$ by 532-nm laser pulses \cite{Polarization}. 
After polarization, the initial state is prepared by the RF $\pi/2$ pulse $R_{\phi}(\pi/2)$, where the rotation axis lies in the XY plane and $\phi =$atan$[(\eta_0^2-1)/2\eta_0]+\pi/2$ is the angle between the rotation axis and the $x$ axis. 
Here we have chosen $\eta(0) = \eta_0I$. 
The evolution under $H_{\rm tot}$ is realized by applying two selective microwave (MW) pulses, and the coherent evolution should be long enough to drive the system to a steady state. 
To this end, $H_s$ is multiplied by an overall coefficient $\lambda$ to speed up the evolution and preserve the coherence. 
The strength of the MW pulses is proportional to the value of $\lambda$. 
However, too strong MW pulses may cause strong crosstalk between different subspaces. 
One possible solution is to use a strongly coupled $^{13}\text{C}$ nuclear spin. 
For the NV centers in diamond with $^{12}$C natural abundance, if one chooses $\lambda$ to maintain coherence, the system cannot reach the steady state because of crosstalk. 
Or if one overcomes the crosstalk by using weak MW pulses, the required evolution time is so long that the coherence will greatly decrease during evolution (see Appendix C). 
Therefore, it is challenging to study the knot topological features of NH systems with a $^{12}$C natural abundance diamond. 
To address this issue, we synthesized a diamond with 99.999\% $^{\text{12}}$C isotope abundance by the chemical vapor deposition method. 
With this sample, the dephasing time of the NV center utilized in our experiment is $T_2^{\star} = 78(7)$ $\upmu$s, as shown in Fig.~\ref{Fig2}(c). 
Such a long coherence time enables us to realize evolution under the NH Hamiltonian during which the coherence is preserved and the crosstalk is suppressed. 
To realize the evolution under $H_{\rm tot}$, we choose a proper rotating frame and use two selective microwave sequences with time-dependent amplitude and phase (see Appendix B). 
The eigenvalues are extracted from the population information by setting $R_i = I$ in the measurement sequence. 
The eigenstates of the NH Hamiltonian are reconstructed by applying $R_i = I, R_{-y}(\pi/2)$ and $R_{-x}(\pi/2)$ to the final state of evolution. 
Before readout of the photoluminescence ratio, we apply an RF $\pi/2$ pulse, which changes the basis of the ancilla from $\ket{\pm}_n$ to $\ket{0,1}_n$. 
Finally, selective $\pi$ pulses are applied to reverse the population of electrons or nuclear spin to get a set of equations that relate photoluminescence intensities to populations of each level.
Solving this equation, the populations are obtained.

Both eigenvalues and eigenstates can be obtained from the measurement sequences. 
When $R_i = I$, we obtain a set of populations at different times by varying the evolution time under $H_{\rm tot}$. 
Renormalizing the population as $P_{1}=P_{\ket{1}_e\ket{1}_n}/(P_{\ket{1}_e\ket{1}_n}+P_{\ket{0}_e\ket{1}_n})$ gives the evolution of $P_{1}$ under the NH Hamiltonian. 
The corresponding quasi-momentum $k$ can be extracted by fitting $P_{1}$ under fixed model parameters. 
Figure~\ref{Fig2}(e) shows the result for the Hopf link phase with $k = 0.6\pi$. 
The population evolution agrees with the theoretical prediction and gives $k_{\rm fit} = 0.59(7)\pi$. 
The eigenvalues can thus be computed from $k_{\rm fit}$ and model parameters. 
By measuring the populations under different bases, we obtain the expectation $\braket{\sigma_{x,y,z}}$. 
Then we can reconstruct the steady state by $\rho=(I+\braket{\vec{\sigma}}\vec{\sigma})/2$, but the direct result may give a mixed state or even an unphysical state. 
Thus, we use the maximum likelihood estimation (MLE) to obtain a pure state close to the eigenstate (see Appendix D). 
Figure~\ref{Fig2}(f) shows the real and imaginary parts of the measured state and theoretical results for $k=0.6\pi$ in the Hopf link phase. 
Here $\rho_1=\ket{\psi_1}\bra{\psi_1}$ is the density matrix corresponding to $\ket{\psi_1}$. The fidelity is 99\% for this state. 
All other states are obtained by following the same procedure and the fidelities are all higher than 97\%.

\begin{figure}[http]
	\centering
	\includegraphics[width=1\columnwidth]{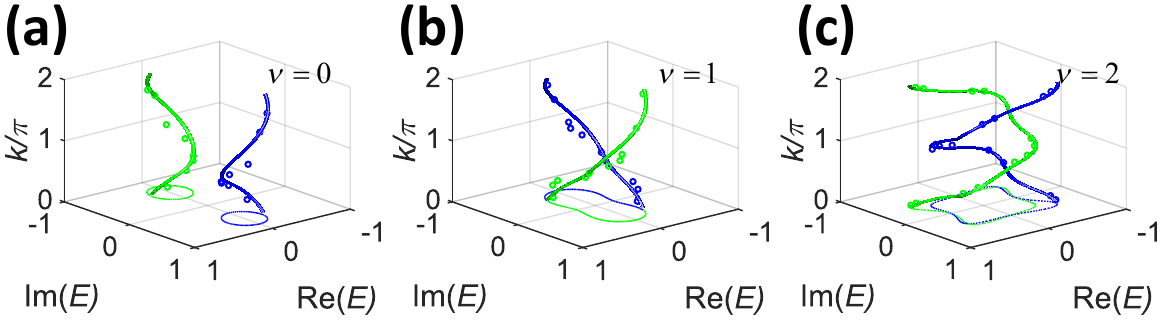}
	\caption{Experimental results of energy eigenvalues for different phases. Red and blue points are experimental results for each band. Lines correspond to theoretical predictions. (a) Unlink phase with wingding number $\nu=0$. (b) Unknot phase with $\nu = 1$. (c) Hopf link phase with $\nu = 2$.}
	\label{Fig3}
\end{figure}

The experimentally measured eigenvalues of different phases for various $k$ are plotted in Fig.~\ref{Fig3}. 
The $B(2)$ braiding behavior is characterized by the following definition of the winding number \cite{Nature_Fan}:
\begin{equation}
	\nu = \int_{0}^{2\pi} \frac{dk}{2\pi i}\frac{d}{dk}\text{lnDet}\left\lbrace H^{(m)}(k)-\frac{1}{2}\text{Tr}[H^{(m)}(k)]\right\rbrace .
	\label{WindingNum}
\end{equation}
This number $\nu$ reflects how many times the energy bands are braided. 
The term $\text{Tr}[H^{(m)}(k)]/2$ here can eliminate the dependence on the choice of reference energy \cite{Nature_Fan}. 
So, the winding number defined in Eq.~(\ref{WindingNum}) only captures the mutual braiding of the eigenvalues under investigation.
Every element of $B(2)$ can be expressed as $\tau^n$, where $n$ is an integer and $\tau$ is the generator. 
The one-to-one correspondence of $\nu=n$ exactly describes the $B(2)$ behavior. 
When $m = 1, \Gamma^{0-} = -0.45, \Gamma^{0+}=0.79, \Gamma^{1-}_{1}=-0.30i, \Gamma^{1-}_{2}=0.08i,$ and $\Gamma^{1+}_{1,2}=0$, the system is in the unlink phase (Fig.~\ref{Fig3}(a)). 
The eigenvalues form two separated curves and $\nu=0$ in this phase. 
The two circles can be separated by a line, $\text{Im}(E)=0$ for example. 
This is just like a normal insulator in the Hermitian case. 
If we set the parameters as $\Gamma^{0-} = -0.21$ and $\Gamma^{0+}=0.70$, the system will be in the unknot phase, as shown by Fig.~\ref{Fig3}(b). 
The two bands interchange as $k$ goes from $0$ to $2\pi$ and in this phase $\nu = 1$. 
Note that $k=0$ and $2\pi$ are equivalent points and in this phase $E_1(0) = E_2(2\pi)$. 
Thus two energy bands form a whole circle instead of two circles. 
For $m = 2$, and $\Gamma^{0-} = 0.04, \Gamma^{0+}=0.49, \Gamma^{1-}_{1}=\Gamma^{1+}_{2}=-0.13i, \Gamma^{1-}_{2}=\Gamma^{1+}_{1}=0.02i, \Gamma^{2-}_{1}=-0.58i, \Gamma^{2+}_{2}=-0.21i, \Gamma^{2+}_{1}=0.03i,$ and $\Gamma^{2-}_{2}=0.09i$, the eigenvalues form a non-trivial braiding pattern (Fig.~\ref{Fig3}(c)) and in this phase $\nu = 2$. 
The two bands encircle each other and form a structure called the Hopf link. 
Although in this phase each band forms a circle, they cannot be separated by any line in the complex energy plane as done in the unlink phase.

\begin{figure}[http]
	\centering
	\includegraphics[width=1\columnwidth]{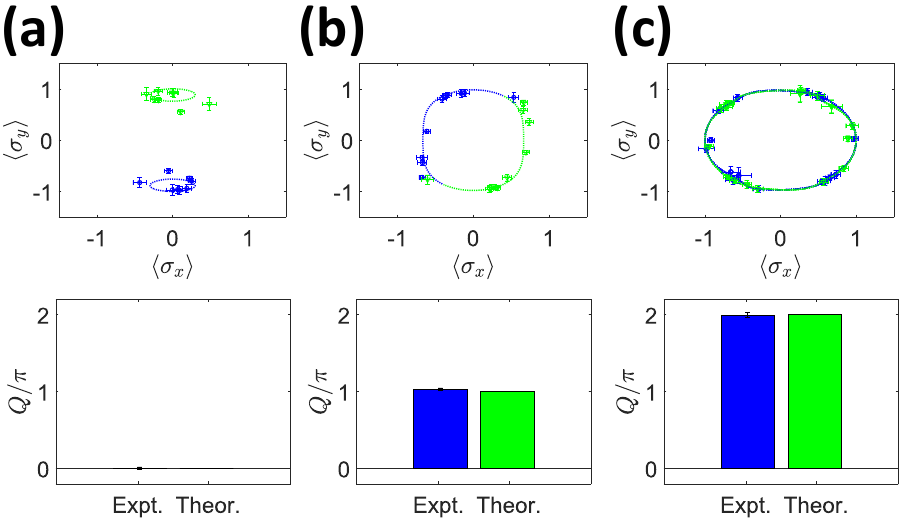}
	\caption{Experimental results of the eigenstates projected on the XY plane and the global Berry phase for each phase. Dashed lines are theoretical predictions. (a) The eigenstates are separated in two regions for the unlink phase. The measured global Berry phase is $Q = 0.00(2)\pi$. (b) The eigenstates form a circle for the unknot phase with  $Q = 1.03(2)\pi$. (c) Each eigenstate forms a closed loop, and the two loops intersect each other for the Hopf link phase. The Berry phase is measured to be $Q = 2.00(3)\pi$.}
	\label{Fig4}
\end{figure}

Apart from the non-trivial topology of the energy bands, the global Berry phase determined by the eigenstates was also observed. 
The global Berry phase is $Q = \int_{0}^{2\pi} $Tr$[A(k)]dk$, where the non-abelian Berry connection is defined as $A_{mn}(k) = i\braket{\chi_m(k)|\partial_k|\psi_n(k)}$ \cite{Q_Liang}. 
Here $\ket{\psi_n(k)}$ ($\ket{\chi_m(k)}$) are right (left) eigenstates of an NH Hamiltonian and satisfy the biorthogonal relation $\braket{\chi_m(k)|\psi_n(k)} = \delta_{mn}$. 
The global Berry phase $Q$ can identify topological invariance in our model \cite{Q_Liang}.
For the unlink, unknot and Hopf link phases, $Q_{\rm ideal} = 0, \pi$ and $2\pi$, respectively. 
Figure~\ref{Fig4} shows the measured eigenstates projected on the XY plane and they show the same behavior as the eigenvalues. 
In the unlink phase, $\ket{\psi_{1,2}(k)}$ form two separated circles on the Bloch sphere, and so are the projections on the XY plane. 
In the unknot phase, the eigenstates form an end-to-end circle instead, just as the energy bands exchange each other.
While in the Hopf link phase, each eigenstates form a closed loop, and the two loops intersect each other.
After obtaining $\ket{\psi_{1,2}(k)}$ from MLE, $\ket{\chi_{1,2}(k)}$ can be solved from the biorthogonal relation. 
The experimental value of $Q$ can be obtained with the method in Ref.~\cite{ComputeQ} via $Q = \sum_{i,n}\text{Im}D_i^n$, where $D_i^n = \text{ln}\braket{\chi_n(k_{i+1})|\psi_n(k_i)}$ and $n=1$ and $2$ is the band index. 
The experimental results of $Q$ for the unlink, unknot and Hopf link phases are 0.00(2)$\pi$, 1.03(2)$\pi$ and 2.00(3)$\pi$, respectively, which agree well with theoretical predictions. 
Generally speaking, for $n$ bands models with eigenvalues $\lbrace E_i\rbrace$ and starting at $k=0$, the ordered set $(E_1,E_2,...,E_n)$ goes to $(E_{\sigma(1)},E_{\sigma(2)},...,E_{\sigma(n)})$ as $k$ varies to $2\pi$. 
Then we have $e^{iQ}=(-1)^{P(\sigma)}$, where $P(\sigma)$ is the parity of the permutation. 
For our case, both the unlink phase and the Hopf link phase have even parity since each band returns to itself as $k$ goes from $0$ to $2\pi$. 
Thus in these two phases $Q=0~($mod $2\pi)$. 
Since the energy bands of the unknot phase exchange each other, the parity is odd with $Q=\pi~($mod $2\pi)$.

In conclusion, we have experimentally investigated the knot topology in an 1D NH model based on both eigenvalues and eigenstates. 
The knot structures of eigenvalues, including the unlink, unknot and Hopf link phases, were successfully observed, which manifest the B(2) braid group behavior.
The global Berry phase was measured via high fidelity eigenstates, which serves as the knot invariant to identify the parity of band braiding. 
Our work makes NV center a desirable platform for investigating important non-Hermitian topology. 
The universality of our dilation method for arbitrary dimensional cases and the ground-state three-level structure of the NV center make it possible to explore the knot topology of three-band models. 
For 1D models with three bands, the knotted topological phases are described by the conjugate classes of $B(3)$ \cite{PRL_Haiping,NatPhys_Bouhon}, which host richer topological behaviors since $B(N)$ is not commutative when $N>2$. 
By introducing more momentum space parameters such as $\vec{k}=(k_x,k_y,k_z)$, our platform can be utilized to investigate the knot topology of higher dimensional NH models \cite{PRB_Fan,PRB_ZhiLi}.

This work was supported by the National Key R\&D Program of China (Grant Nos. 2018YFA0306600 and 2016YFB0501603), the National Natural Science Foundation of China (Grant No. 12174373), the Chinese Academy of Sciences (Grant Nos. XDC07000000 and GJJSTD20200001), Innovation Program for Quantum Science and Technology (Grant No. 2021ZD0302200), Anhui Initiative in Quantum Information Technologies (Grant No. AHY050000) and Hefei Comprehensive National Science Center. X. R. thanks the Youth Innovation Promotion Association of Chinese Academy of Sciences for their support. Ya Wang and Yang Wu thanks the Fundamental Research Funds for the Central Universities for their support.

%
Yang Wu, Yunhan Wang and Xiangyu Ye contributed equally to this work. 

\section{Appendix A: Experimental Setup}
\begin{figure}[http]
	
	\centering
	
	\includegraphics[width=1\columnwidth]{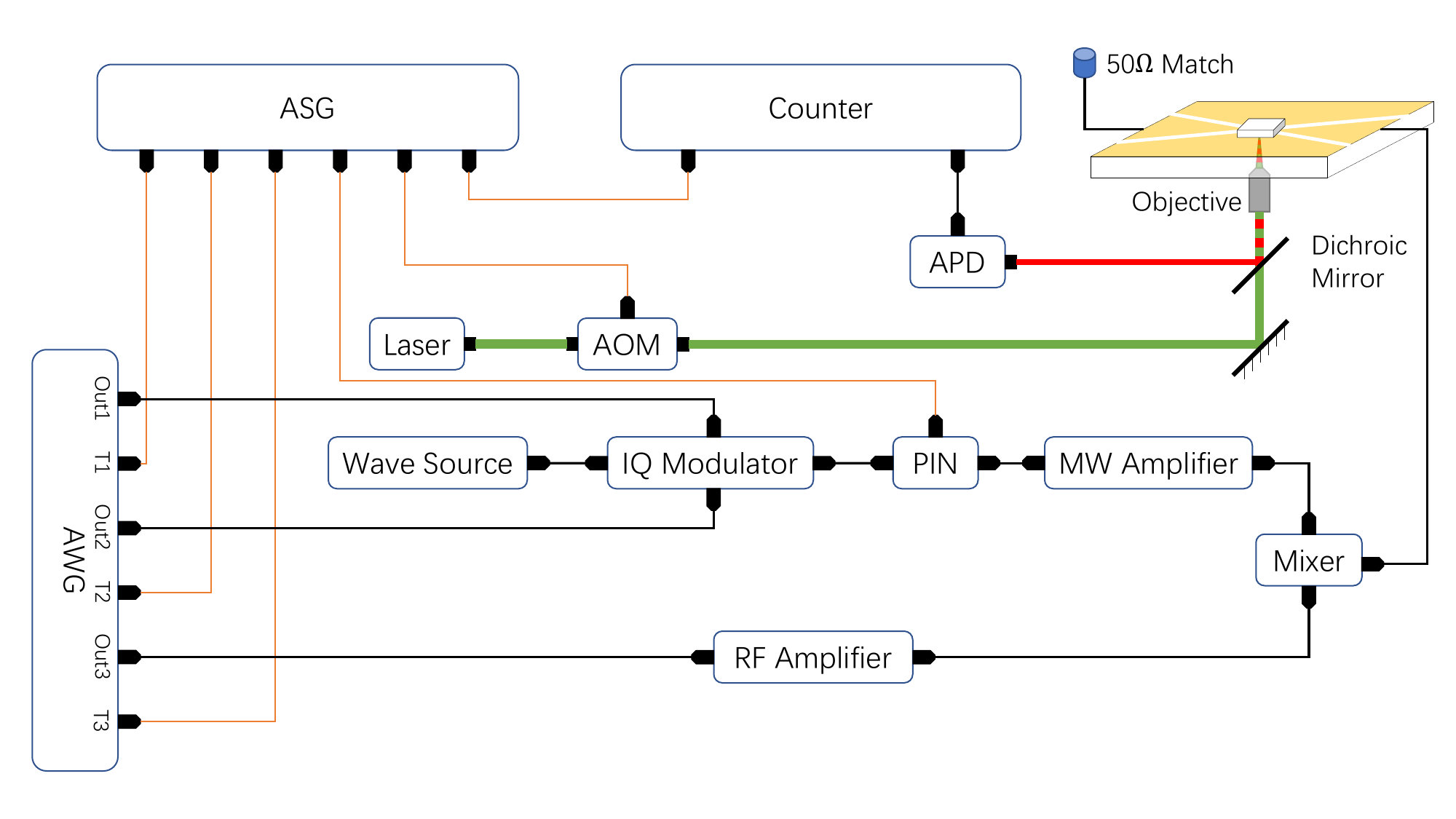}
	\caption{The hardware setup in our experiment. Both MW and RF pulses were generated by the AWG and an IQ modulator was utilized to adjust the frequency of MW pulses to the transition frequency we needed. The pulses were fed by the same coplanar waveguide after being amplified by amplifiers. The 532 nm laser was used to polarize and readout the spin state of the NV center. The on/off of the laser beam was controlled by an AOM. The fluorescence (650-800 nm) from the NV center passed the objective and was collected by an APD. The timing for our experiment was controlled by the arbitrary sequence generator (ASG).
	}
	\label{HWC}
\end{figure}

The experimental setup is shown in Fig.\ref{HWC}. The diamond was mounted on a confocal setup, and the static magnetic field of 506 G was provided by a permanent magnet along the NV symmetry axis. The initialization and the readout of the NV center spin were realized with a 532 nm green laser controlled by an acousto-optic modulator (AOM) (ISOMET, AOMO 3200-121). The laser beam traveled twice through the acousto-optic modulator before going through an oil objective (Olympus, PLAPON 60*O, NA 1.42) and then focusing on the NV center. The phonon sideband fluorescence (wavelength, 650-800nm) went through the same oil objective and was collected by an avalanche photodiode (Perkin Elmer, SPCM-AQRH-14) with a counter card (NI, PCIe-6612).

The radio-frequency (RF) pulses were generated by an arbitrary waveform generator (AWG) (CIQTEK AWG4100) and were amplified by a power amplifier (Mini-Circuits, LZY-22+). The microwave (MW) pulses were generated by the same arbitrary waveform generator. The bandwidth of the arbitrary waveform generator is 0-330 MHz, which is far from the resonant frequency (about 1.47 GHz). Thus, the pulses were mixed with continuous 1.6-GHz output from a wave source (RIGOL, DSG3065B) using an IQ modulator (Marki, IQ1545LMP). Then the pulses passed a PIN (Mini-Circuits, ZASWA-2-50DRA+) and an amplifier (Mini-Circuits, ZHL-15W-422-S+). Finally both the MW and RF pulses were fed by the same coplanar waveguide after passing the diplexer (Marki, DPX-0R5). The waveforms of MW and RF pulses were prepared in advance and were downloaded to the AWG. An arbitrary sequence generator (CIQTEK ASG8100) was utilized to control the timing in experiment by the trigger signals  designed in advance.

\section{Appendix B: Universal Dilation Method}
We use the universal dilation method to construct the NH Hamiltonian \cite{Science_YW}. Intuitively, by introducing an ancilla system and tuning their interaction in a time-dependent way, the target system obeys the evolution governed by the NH Hamiltonian $H_s$. The detail can be found in Ref.~\cite{Science_YW}. For an NH Hamiltonian $H_s$, the dilated Hamiltonian $H_{s,a}$ takes the form:
\begin{equation}
	H_{s,a} = \Lambda(t) \otimes I + \Gamma(t) \otimes \sigma_z,
\end{equation}
where $I$ is the identity matrix and $\sigma_z$ is the Pauli operator on the ancilla. $\Lambda(t)$ and $\Gamma(t)$ are operators on the system written as: $\Gamma(t) = \{ H_s(t) + [i\frac{d}{dt}\eta(t)+\eta(t)H_s(t)]\eta(t)\}M^{-1}(t)$ and $\Lambda(t) = i[H_s(t)\eta(t)-\eta(t)H_s(t)-i\frac{d}{dt}\eta(t)]M^{-1}(t)$. Here $M(t) = \eta(t)^\dagger\eta(t) + I $ and $M(t)$ satisfies the following equation:
\begin{equation}
	i\frac{d}{dt}M(t) =H_s^\dagger(t)M(t)-M(t)H_s(t).
\end{equation}

Now we focus on the realization of $H_{s,a}$ in the NV center.
The ground state of the NV center can be described by the Hamiltonian
\begin{equation}
	H_{NV}=2\pi (D S_z^2+\omega_e S_z+Q I_z^2 + \omega_n I_z + A S_z I_z),
\end{equation}
where $S_z\ (I_z)$ is the spin-1 operator of the electron (nuclear) spin, $D = 2.87$ GHz is the zero field splitting for the electron, $\omega_e\ (\omega_n)$ is the electron (nuclear) Zeeman term induced by the magnetic field applied along the NV symmetry axis, $Q = -4.95$ MHz is the nuclear quadrupolar interaction, and $A = -2.16$ MHz is the hyperfine interaction. We construct the dilated Hamiltonian $H_{\rm tot}$ in the subspace spanned by the states $\ket{m_S,m_I}=\ket{0,1},\ket{0,0},\ket{-1,1}$, and $\ket{-1,0}$. The Hamiltonian in this subspace can be simplified to
\begin{equation}
	H_0 = \pi[-(D-\omega_e-\frac{A}{2})\sigma_z\otimes I+(Q+\omega_n-\frac{A}{2})I\otimes\sigma_z+\frac{A}{2}\sigma_z\otimes\sigma_z].
\end{equation}
To facilitate the construction of $H_{s,a}(t)$, we further decompose $H_{s,a}(t)$ as:
\begin{equation}
	\begin{split}
		H_{s,a} &= B_1 I\otimes I + A_1 \sigma_x\otimes I + B_2 \sigma_y\otimes I+ B_3\sigma_z\otimes I\\
		&+A_2I \otimes\sigma_z  + B_4 \sigma_x \otimes \sigma_z + A_3\sigma_y \otimes \sigma_z + A_4 \sigma_z \otimes \sigma_z,
	\end{split}	
\end{equation}
where $A_i$ and $B_i$ are time-dependent coefficients. We apply selective MW pulses to cast the control Hamiltonian:
\begin{equation}
	\begin{split}
		H_c(t)&=2\pi\Omega_1(t)\cos[\int_{0}^{t}\omega_1(\tau)d\tau+\phi_1(t)]\sigma_x\otimes \ket{1}_n\bra{1}\\
		&+2\pi\Omega_2(t)\cos[\int_{0}^{t}\omega_2(\tau)d\tau+\phi_2(t)]\sigma_x\otimes \ket{0}_n\bra{0}.
	\end{split}
\end{equation}
In order to realize $H_{s,a}$, we choose the rotating frame with the following form:
\begin{equation}
	U_{rot} = e^{i\int_{0}^{t}H_0-B_1(\tau)I\otimes I-B_3(\tau)\sigma_z\otimes I-A_2(\tau) I\otimes\sigma_z-A_4(\tau)\sigma_z\otimes\sigma_zd\tau}.
\end{equation}
Thus the Hamiltonian in the rotating frame can be written as
\begin{equation}
	\begin{split}
		&H_{tot} = U_{rot}(H_0+H_c)U_{rot}^\dagger-iU_{rot}\frac{dU_{rot}^\dagger}{dt}\\
		&= B_1I\otimes I+B_3\sigma_z\otimes I+A_2I\otimes\sigma_z+A_4\sigma_z\otimes\sigma_z\\
		&+2\pi\Omega_1\cos(\phi_1+\int_{0}^{t}\omega_1)\cos(\int_{0}^{t}\tilde{\omega}_1+2B_3+2A_4)\sigma_x\otimes\ket{1}_n\bra{1}\\
		&+2\pi\Omega_1\cos(\phi_1+\int_{0}^{t}\omega_1)\sin(\int_{0}^{t}\tilde{\omega}_1+2B_3+2A_4)\sigma_y\otimes\ket{1}_n\bra{1}\\
		&+2\pi\Omega_2\cos(\phi_2+\int_{0}^{t}\omega_2)\cos(\int_{0}^{t}\tilde{\omega}_2+2B_3-2A_4)\sigma_x\otimes\ket{0}_n\bra{0}\\
		&+2\pi\Omega_2\cos(\phi_2+\int_{0}^{t}\omega_2)\sin(\int_{0}^{t}\tilde{\omega}_2+2B_3-2A_4)\sigma_y\otimes\ket{0}_n\bra{0}.
	\end{split}
	\label{no_cross}
\end{equation}
We have omitted the $t$ dependence and the integral variable for simplicity, and $\tilde{\omega}_{1}$ $(\tilde{\omega}_2)$ is the transition frequency of between $\ket{0,1}$ and $\ket{-1,1}$ ($\ket{0,0}$ and $\ket{-1,0}$). In order to reduce $H_{tot}$ to $H_{s,a}$ under the rotating-wave approximation, the amplitudes, frequencies and phases should satisfy $\omega_1=\tilde{\omega}_1+2B_3+2A_4,\omega_2=\tilde{\omega}_2+2B_3-2A_4$ and
\begin{equation}
	\left\{
	\begin{aligned}
		&\frac{\pi}{2}(\Omega_1\cos\phi_1+\Omega_2\cos\phi_2)=A_1,\\
		&\frac{\pi}{2}(\Omega_1\cos\phi_1-\Omega_2\cos\phi_2)=B_4,\\
		&\frac{\pi}{2}(-\Omega_1\sin\phi_1-\Omega_2\sin\phi_2)=B_2,\\
		&\frac{\pi}{2}(-\Omega_1\sin\phi_1+\Omega_2\sin\phi_2)=A_3.
	\end{aligned}
	\right.
\end{equation}
The solution reads
\begin{equation}
	\left\{
	\begin{aligned}
		\Omega_1&=\frac{\sqrt{(A_1+B_4)^2+(B_2+A_3)^2}}{\pi},\\
		\Omega_2&=\frac{\sqrt{(A_1-B_4)^2+(-B_2+A_3)^2}}{\pi},\\
		\phi_1&=\text{atan2}(-B_2-A_3,A_1+B_4),\\
		\phi_2&=\text{atan2}(A_3-B_2,A_1-B_4).
	\end{aligned}
	\right.
\end{equation}

\begin{figure}[http]
	\centering
	\includegraphics[width=0.8\columnwidth]{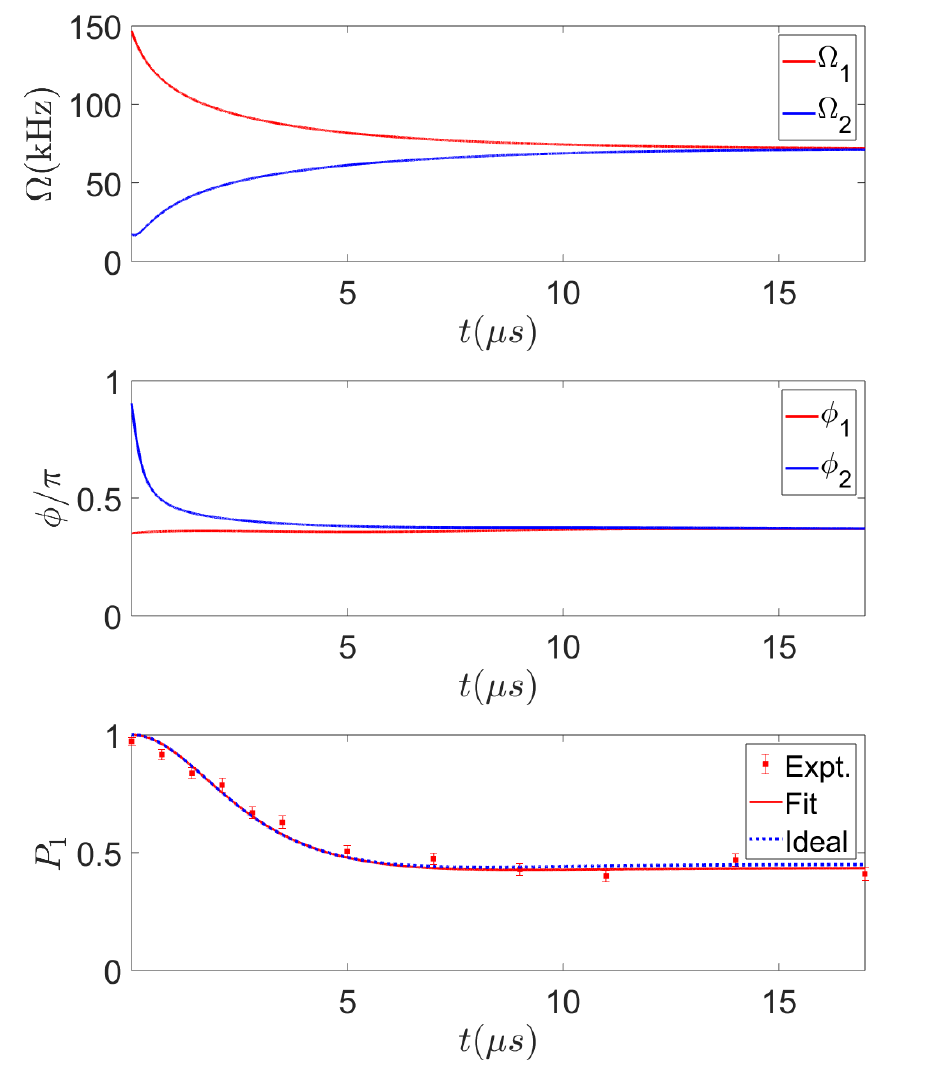}
	\caption{The time-dependent amplitudes and phases of the control pulses and the corresponding experimental result of the population evolution. The results correspond to the Hopf link phase with $k = 1.65\pi$.}
	\label{SFig5}
\end{figure}

As an example, the result for the Hopf link phase with $k = 1.65\pi$ is shown in Fig.~\ref{SFig5}. The target NH Hamiltonian is 
\begin{equation}
	\begin{aligned}
		&H^{(2)}(k) = \left[\begin{array}{ccc}
			0 & \Gamma^{0-}\\
			\Gamma^{0+} & 0
		\end{array}\right] + \\
		&\sum_{n=1}^{2}\left[\begin{array}{ccc}
			0 & \Gamma_{1}^{n-}e^{ink}+\Gamma_{2}^{n+}e^{-ink}\\
			\Gamma_{1}^{n+}e^{-ink}+\Gamma_{2}^{n-}e^{ink} & 0
		\end{array}\right],
	\end{aligned}
\end{equation}
where $\Gamma^{0-} = 0.04, \Gamma^{0+} = 0.49, \Gamma_{1}^{1-} = \Gamma_{2}^{1+}= -0.13i$, $\Gamma_{2}^{1-} =\Gamma_{1}^{1+} = 0.02i, \Gamma_{1}^{2-} = -0.58i, \Gamma_{2}^{2+} = -0.21i$, $\Gamma_{1}^{2+} = 0.03i,  \Gamma_{2}^{2-} = 0.09i$, and $k = 1.65\pi.$ Since the eigenvalues of the Hamiltonian are complex, the state evolution shows a damping behavior and finally reaches one of the eigenstate of the NH Hamiltonian $H_s$. The dilation procedure gives $\Omega_i(t)$ and $\phi_i(t)$ to realize the dilated Hamiltonian $H_{tot}$, as shown in Fig.~\ref{SFig5}. 
For an intuitive interpretation, both $\Omega_i(t)$ and $\phi_i(t)$ first show a decay behavior and then almost remain unchanged. When $\Omega_i(t)$ and $\phi_i(t)$ are almost equal to their steady value, they effectively act as a rotation along a specific axis determined by the eigenstate of the NH Hamiltonian in the target subspace. The beginning decay part of the control Hamiltonian exactly drives the initial state to be parallel with this rotation axis. The final state thus remains unchanged (in the target subspace) though $\Omega_i$ is not zero. The measured population evolution agrees well with this picture and the theoretical predictions.
\begin{figure}[http]
	\centering
	\includegraphics[width=1.5\columnwidth]{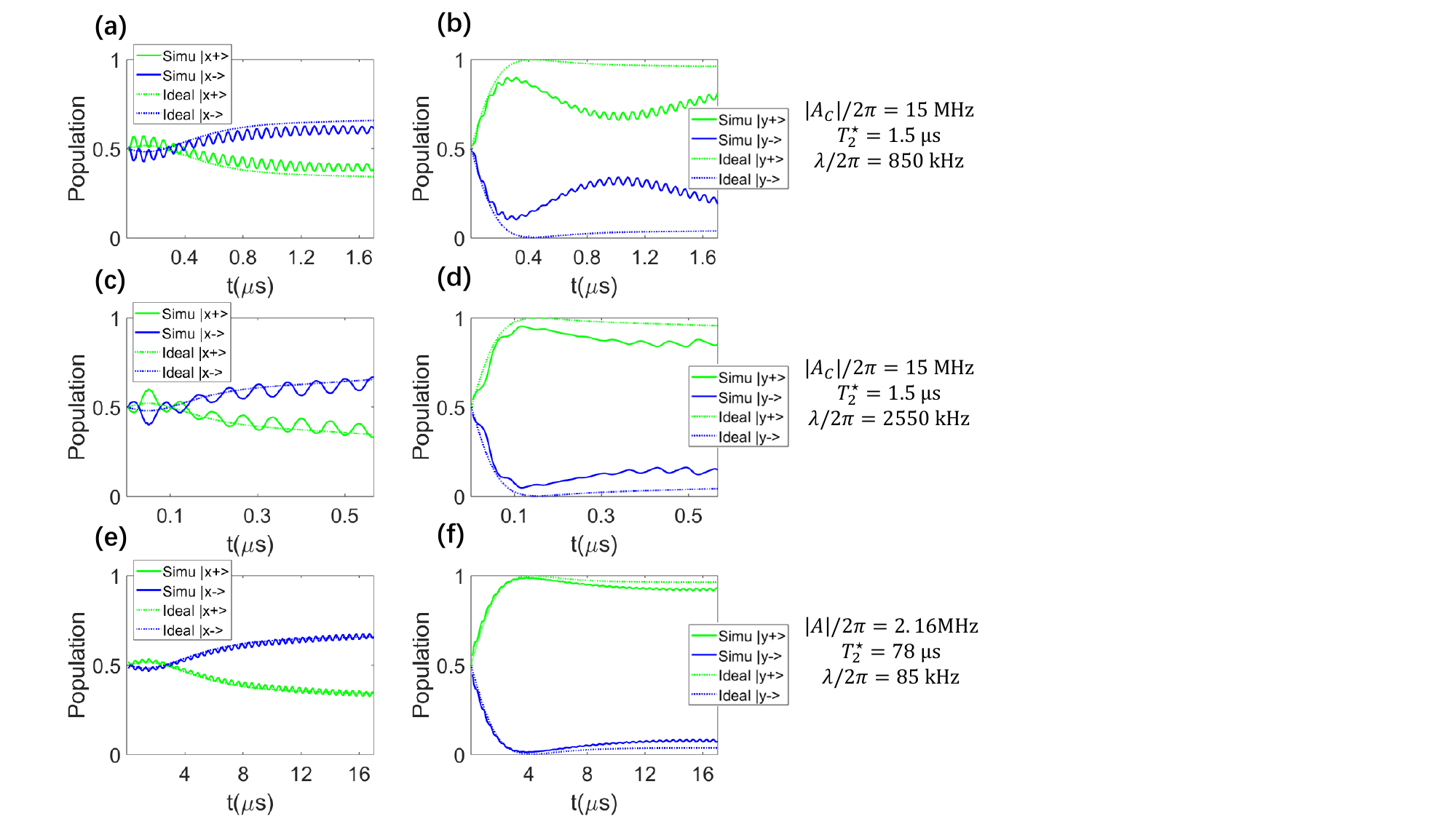}
	\caption{Simulation results of population evolution in the $\sigma_x$ (a,c,e) basis and the $\sigma_y$ (b,d,f) basis. (a-d) NV centers in diamond with $^{12}$C natural abundance and different $\lambda$'s. (e,f) Our parameters and NV center in $^{12}$C purified diamond. The parameters are as follows: (a,b) $\lambda=2\pi\times850$ kHz, $T_2^\star = 1.5\ \upmu$s; (c,d) $\lambda=2\pi\times2550$ kHz, $T_2^\star = 1.5\ \upmu$s; and (e,f) $A = -2.16$ MHz, $T_2^\star = 78\ \upmu$s, and $\lambda=2\pi\times85$ kHz.
	}
	\label{SFig1}
\end{figure}

\section{Appendix C: Effects of Dephasing and Crosstalk}
The evolution under the NH Hamiltonian is mainly affected by the dephasing and the crosstalk. Equation~\ref{no_cross} shows the ideal case when the two MW pulses are selective. However, in practice the operators in the control Hamiltonian $H_c$ have the form $\sigma_x\otimes I$ instead of $\sigma_x\otimes\ket{1}_n\bra{1}$, and the decoherence inevitably undermines the evolution. As mentioned in the main text, one way is to use a strongly coupled $^{13}$C nuclear spin. For a natural abundance diamond, the dephasing time $T_2^\star$ ranges from $1 \ \upmu$s to $3\ \upmu$s. We take $T_2^\star =1.5\ \upmu$s as a typical value. The coupling strength $A_C$ between the $^{13}$C nuclear spin and the electron spin is on the order of 10 MHz generally and we take $|A_C| = 15$ MHz. The simulation results of the evolution under the $\sigma_x$ basis for the Hopf link phase with $k=0.85\pi$ and different $\lambda$'s are shown in Fig.~\ref{SFig1}. If one uses weak driving to suppress the crosstalk, the evolution time is $1.5\ \upmu$s, which is comparable to the coherence time. The decoherence drastically destroys the coherent evolution. The population evolution under $\sigma_{x,y}$ bases is shown in Figs.~\ref{SFig1}(a) and \ref{SFig1} (b), which greatly deviates from the ideal case. We use $c=\braket{\sigma_x}^2+\braket{\sigma_y}^2$ to characterize the coherence. In this parameter configuration, for the final state we have $c=0.65$ and for the ideal case we have $c=0.97$. Only about 67\% coherence is left. One may try to set a large value of $\lambda$ under which the time needed to reach the steady state is within the coherence time. When $\lambda/2\pi$ varies from $850$ kHz to $2550$ kHz, $\lambda/|A_C|$ increases from 0.057 to 0.17 and the evolution time needed decreases from $1.7\ \upmu$s to $0.6\ \upmu$s. The pulses are no longer selective and the evolution is significantly affected by the crosstalk. As can be seen from Fig.~\ref{SFig1}(c) and \ref{SFig1}(d), the system can barely reach a steady state.

Another way is to use samples in which the NV center has a long coherence time. For our sample, fitting of the luminescence in the Ramsey experiment shows that $T_2^\star = 78(7)\ \upmu$s (see main text). Fig.~\ref{SFig1}(e) and \ref{SFig1}(f) show the simulation results for our parameters where we take $T_2^\star = 78\ \upmu$s. Here $A=-2.16$ MHz is a typical value for the coupling strength between the $^{14}$N nuclear spin and the electron spin. For this parameter configuration, $\lambda/|A|=0.039$. The deviation caused by decoherence and crosstalk is about 0.03 and these effects on the fidelity between the final states in experiments and the ideal eigenstates can be ignored.

\section{Appendix D: Experimental Acquisition of the Eigenstates}

\begin{figure}[http]
	
	\centering
	
	\includegraphics[width=2.2\columnwidth]{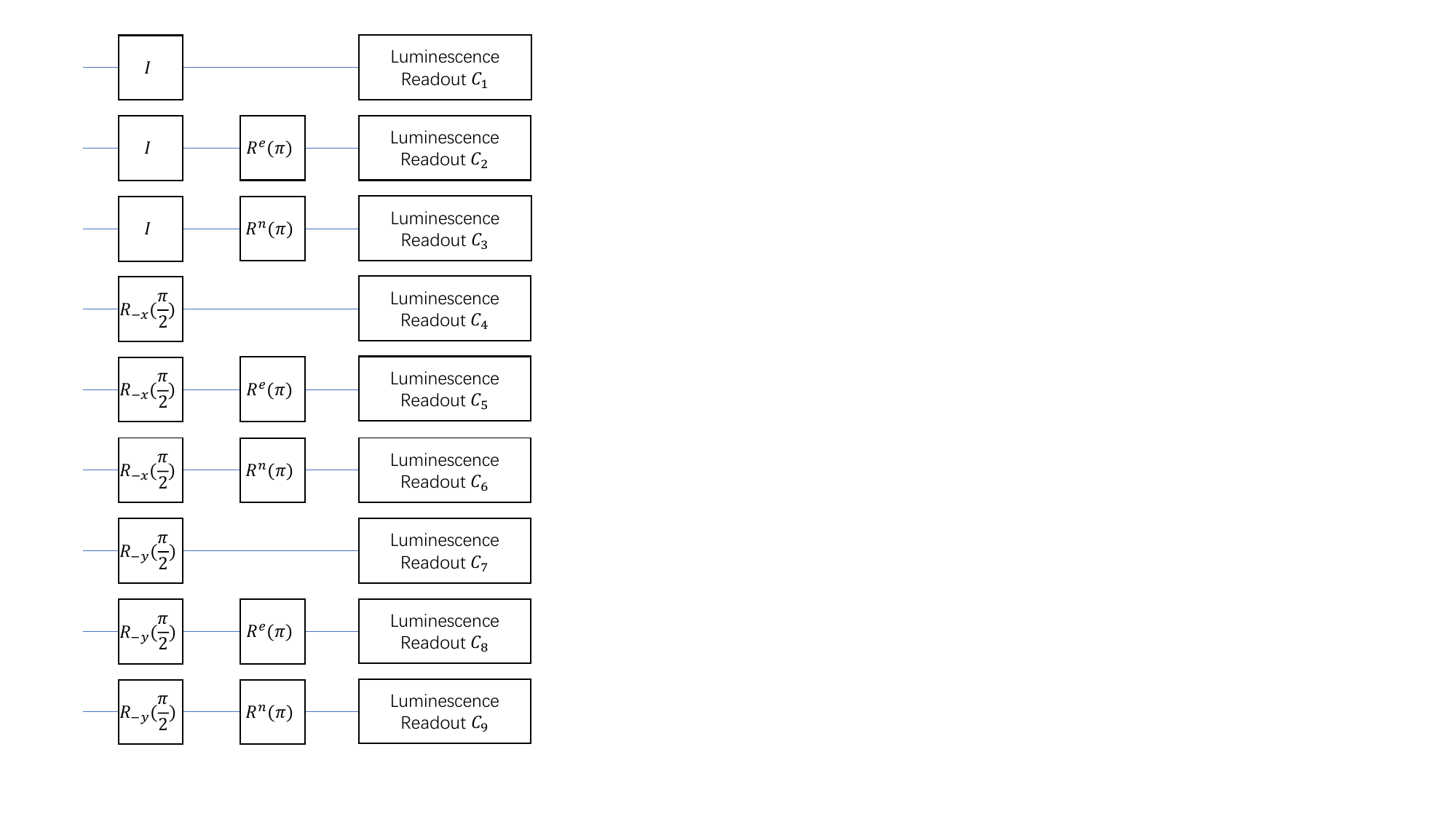}
	\caption{Measurement sequences to reconstruct the state. Counts $C_1\sim C_9$ are measured counts for each sequence.
	}
	\label{SFig2}
\end{figure}

For the final state of the evolution, nine different measurement sequences are applied to reconstruct the state (Fig.~\ref{SFig2}). Here $R^e(\pi)$ is the $\pi$ pulse on the electron spin in the $\ket{1}_n$ subspace and $R^n(\pi)$ is the $\pi$ pulse on the nuclear spin in the $\ket{0}_e$ subspace. $R^e(\pi)$ and $R^n(\pi)$ correspond to the transition MW1 and RF2, respectively, in Fig.2(d) of the main text. The $\pi/2$ pulses rotate the electron spin in both spaces. We use normalization sequences to obtain the PL rate for each of the four levels \cite{PRL_Wenquan}. The contribution for the counts of each level is the PL rate multiplied by the corresponding population. The count of each measurement sequence equals the summation of the contribution over each level. By solving the equations that relate the populations of each level under different bases and counts for each measurement sequence, we can obtain the expectation values $\braket{\sigma_{x,y,z}}$ of the final state.

From $\braket{\sigma_{x,y,z}}$ of the final state, we can directly reconstruct $\rho$ by $\rho = (I+\braket{\vec{\sigma}}\vec{\sigma})/2$, but the direct result may give a mixed state or even an unphysical state. Thus the maximum likelihood estimation has been utilized to obtain the pure states from the experimental results. We parameterize the pure state as $(\alpha\ket{0}_e+\beta e^{i\gamma}\ket{-1}_e)\ket{1}_n+(\delta\ket{0}_e+\epsilon e^{i\zeta}\ket{-1}_e)\ket{0}_n$, where all parameters are real and satisfy the normalization condition. Note that the measurement result of each sequence is determined by the population on each level. Since $R^n(\pi)$ and $R^e(\pi)$ reverse the population of the corresponding levels, the phase difference between the two subspace does not manifest in the measurement result. Here we fix the coefficients of $\ket{0}_e$ to be real to eliminate the irrelevant phase freedom, since we only need the results in the $\ket{1}_n$ subspace. Then the expectation values for the nine counts can be obtained from the PL rates $p_{1,2,3,4}$ and the parameters of the pure state, where we label the levels $\ket{m_S,m_I}=\ket{0,1},\ket{-1,1},\ket{0,0}$, and $\ket{-1,0}$ as $1,2,3,$ and $4$ for simplicity. For example, the expectation values for the counts of the first and second sequences read $\tilde{C}_1=\alpha^2p_1+\beta^2p_2+\delta^2p_3+\epsilon^2p_4$ and $\tilde{C}_2=\beta^2p_1+\alpha^2p_2+\delta^2p_3+\epsilon^2p_4$. Then the loss function is chosen to be
\begin{equation}
	L(\alpha,\beta,\gamma,\delta,\epsilon,\zeta)=\sum_{i=1}^{9}(C_i-\tilde{C}_i)^2,
\end{equation}
where $C_i$ are the measured counts for each sequence. Optimize these parameters to minimize $L$ and we obtain the pure state $\alpha_{\text{min}}\ket{0}_e+\beta_{\text{min}} e^{i\gamma_{\text{min}}}\ket{-1}_e$ up to a normalization factor. Here the subscript means the parameters that minimize $L$. 
The experimentally obtained fidelities of all the eigenstates exceed 0.97.
Based on the model parameters given in the main text, we show the results of some eigenstates $\psi_{1,2}$ as examples in Fig.\~ref{SFig3}. 
The corresponding fidelities are as follows: Hopf link, 1.00(7) and 1.00(6) for $\psi_1$ and $\psi_2$ (same below) at $k=0.1\pi$; unknot, 1.00(2) and 1.00(3) at $k=\pi$ and unlink, 1.00(3) and 1.00(3) at $k=0.6\pi$. 

\begin{figure*}[http]
	
	\centering
	
	\includegraphics[width=2.0\columnwidth]{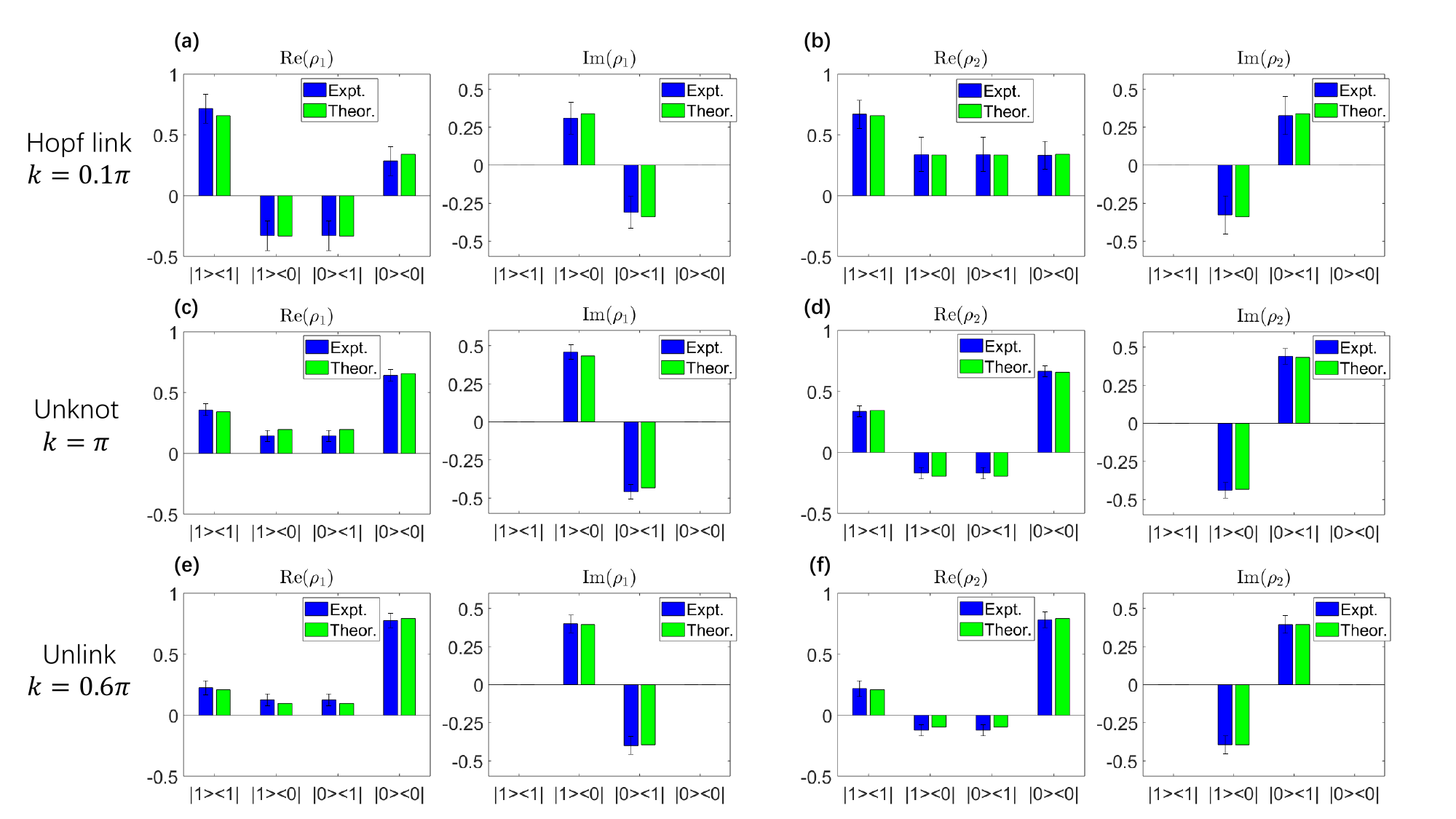}
	\caption{Examples for measured states in each phase. (a,b) The Hopf link phase with $k=0.1\pi$. Panel (a) shows the real and imaginary parts for the one eigenstate, and panel (b) shows the real and imaginary parts of the other eigenstate of the same Hamiltonian. The arrangement of other sub-figures is similar. (c,d) The unknot phase with $k=\pi$. (e,f) The unlink phase with $k=0.6\pi$.
	}
	\label{SFig3}
\end{figure*}

\newpage

%
%

\end{document}